\documentclass[12pt]{article}

\usepackage{amsmath,amssymb}
\usepackage{braket}
\usepackage{graphicx}
\usepackage{color}
\usepackage{dcolumn}
\usepackage{bm}
\usepackage[numbers,super,comma,sort&compress]{natbib}

\usepackage{authblk}
\definecolor{background-color}{gray}{0.98}

\title{Selective hydrogen production at platinum surfaces investigated by the Quantum Monte Carlo approach to chemical reactions.}

\author{Rajesh O. Sharma$^1$, Tapio Rantala$^2$ and Philip E Hoggan$^1$}

\affil{ (1) Institute Pascal, UMR 6602 CNRS,BP 80026, 63177 Aubiere Cedex, France, \\ (2)	Physics, Tampere University, Tampere, Finland.}

\begin{document}

\maketitle

\begin{abstract}
The present letter describes an atomic scale investigation of a chemical reaction for selective hydrogen production. This clean fuel is a sustainable energy source. Because electron transfer is the key to such reactions, accurate methods based on quantum theory are used.
The reaction between water and carbon monoxide has been used industrially with metal catalysts, usually Platinum. There is a considerable economic and environmental challenge underpinning this application of a fundamental process limited by bond dissociation. That is the process often limiting reaction rates for industrial catalysis. Most mainstream quantum approaches fail to greater or lesser degree in the description of bond dissociation.
The present work presents a promising alternative: the initial analysis of a considerable mass of statistical data generated by the atomic-scale Quantum Monte Carlo method to very stringent statistical accuracy for essential information on the hydrogen production via the water-gas shift reaction with platinum catalyst.

This is encouraging for establishing less well-known benchmark values of industrial reaction barriers on Pt(111).
\end{abstract}

Keywords: Quantum Monte Carlo calculation, heterogeneous catalysis, metal surface, low activation barrier

\clearpage
\makeatletter
\renewcommand\@biblabel[1]{#1.}
\makeatother
\bibliographystyle{apsrev}
\renewcommand{\baselinestretch}{1.5}
\normalsize
\clearpage

\section{\label{sec:1}Introduction}

Hydrogen is a clean fuel obtained by catalytic addition of water to carbon monoxide. The reactants are in equilibrium with products and it is energetically less-favored in the direction described here, producing hydrogen. It is known as the water-gas shift (wgs) reaction and is greatly facilitated on a platinum surface. The wgs reaction follows the equation below:

CO + H$_2$O $\rightarrow $  CO$_2$ + H$_2$

In spite of much work on the water-gas shift reaction (e.g. \cite{wgs}), including its use to produce hydrogen, the mechanism and energy barrier are sparsely documented, although various approaches are assembled in \cite{mukin} and the present work aims to provide information obtained by calculation at an atomic-electronic scale.

Stochastic approaches to the Schr{\"o}dinger equation (for systems evolving with time) are currently attracting significant rapidly growing interest. Such an approach is the Quantum Monte Carlo (QMC) method, which is applied here. Errors can be made small, given time and the procedure scales well on highly parallel computers. This work gives so-called ‘chemical accuracy’ for the water gas-shift activation energy barrier (i.e. energy to within 1 kcal/mol).

The hydrogen produced (in the forward wgs) is a sustainable energy source, with combustion giving water which is one of the wgs reactants.

The mechanistic study described in \cite{mukin} assumes water dissociation is rate-limiting. It presents DFT, experiment and micro-kinetic modelling. The experimental apparent activation energy it cites as 17.05 kcal/mol is almost exactly the value obtained here by Quantum Monte Carlo (QMC). Conversely, the DFT evaluation is 16.19 kcal/mol, significantly lower than more recent work using a similar approach \cite{phat,absi}. A comparative DFT study  \cite{phat} gave 17.99 kcal/mol for the water dissociation whereas our QMC estimate for this pathway is 17.7 kcal/mol  \cite{absi}.
We found a preferred concerted mechanism, with a QMC barrier of 17.0 $\pm$ 0.8 kcal/mol. Note also that water dissociation leads to adsorbed OH in \cite{mukin}, which is not reactive towards the pre-adsorbed CO, unlike water with an OH bond stretch that can approach the CO in a suitable mutual orientation (see below). Previously published activation barriers for CO oxidation (involved in water-gas shift reaction mechanisms) are cited in \cite{mukin} over a range covering a factor of two: 11.3 to 23.3 kcal/mol.

Chemical accuracy for the energy barriers encountered by reacting molecules (activation energy) is difficult to obtain. It is determined as the difference between two large and similar energies for the same atoms in different geometries.

1-The asymptotic geometry.

2-The Transition-state geometry.

In such a case, both energies need to be determined very accurately and the second geometry (so-called transition-state, TS) is difficult to locate.

Atoms with substantial interaction during the wgs reaction define its molecular and solid ‘active site’ (see Figue 1, below). This 'active site' system is embedded in a periodic solid, i.e. the metal catalyst.
The metal is platinum, with an exposed compact Pt(111) face (where the atoms are arranged in a honeycomb lattice). Reactant molecule interaction at this surface tips the water-gas shift equilibrium towards hydrogen as a product.
In quantum methodology, there are many well-established methods but few describe bond-dissociation (breaking) well. One method that does deliver for molecule-solid systems, as we show here, is Quantum Monte Carlo (QMC). As the name suggests, this approach is stochastic (or statistical). It is based on the idea of describing many electrons by random-walks, subject to their interactions with each other and with atoms. QMC is not, in fact, restricted to a type of particle, in terms of the algorithm. To describe electrons, users must input a guiding ‘wave-function’ that has electronic anti-symmetry. This guiding wave-function must be as accurate as possible to limit the related systematic error. It follows that embedding a very precise ‘molecular’ wave-function for the molecules and active site into a convenient periodic wave-function for the solid as a whole is a promising approach.

\section{\label{sec:2}Preparing the trial wave-function.}
In this work, the molecular part of the wave-function was prepared using the well-known molecular ab initio software, MOLPRO \cite{molpro}, resorting to Multi-Reference Configuration Interaction (MRCI). The Full Configuration Interaction was carried out with NECI \cite{neci} and the TS geometry determined by QMC. From this large wave-function, the top (most weighted) configurations were retained for embedding in a periodic plane-wave function describing the whole system, including the platinum substrate other that two of its atoms that were included in the active site, one to represent the triangle of equivalent atoms reacting with water at the close-packed (hexagonal in 2D) Pt(111) surface and the other the Pt-atom to which the reacting carbon monoxide molecule was initially bound.
How this reaction proceeds is a matter of intense research currently. It is generally accepted that the carbon monoxide molecule binds first, on top of a platinum atom in the surface. The carbon atom is linked to the metal and becomes the seat of a partial positive charge, which makes it reactive towards oxygen atoms, like that in the water molecule. Afterwards, some authors suggest dissociation of water to form a hydroxyl radical (.OH) and a nascent hydrogen atom. An argument in favour of this would be the reactivity of the OH-radical although \cite{mukin} indicates this reactivity leads to OH-radical adsorption. Whatever reaction these radicals undergo, the O-H bond in water is very costly to break in energy terms and only a part of this energy could be recovered by forming a Pt-H bond at the surface.

\section{\label{sec:3}Preparing locating the transition-state.}
QMC estimates of all force constant (second derivatives with respect to internal co-ordinates) are evaluated to locate a TS. This is the saddle-point on the reaction path, so the force-constant in this direction is negative and all others positive.  The water geometry has a O-H stretched whilst the oxygen is already beginning to bind to the carbon atom of the carbon monoxide molecule. This assumes that these molecules can approach each other, at the Pt(111) surface. We have seen that the carbon monoxide tends to be quite strongly adsorbed onto a given Pt-atom, conversely, the water molecule is mobile and can diffuse at the surface, whilst the hydrogen atom of the stretched O-H bond binds to a neighbouring Pt-atom in this surface.
In the asymptotic geometry, the CO molecule is already bound to a surface Pt-atom and the water molecule, in its equilibrium geometry is a long distance further along the surface (7 \AA ngstoms).

The TS geometry we have located is illustrated below. The geometry may not have a stand-out structure but its energy should be the saddle point value (maximum of the path from reactants to products; like the top of a mountain pass).

\begin{figure}
\includegraphics[scale=0.75]{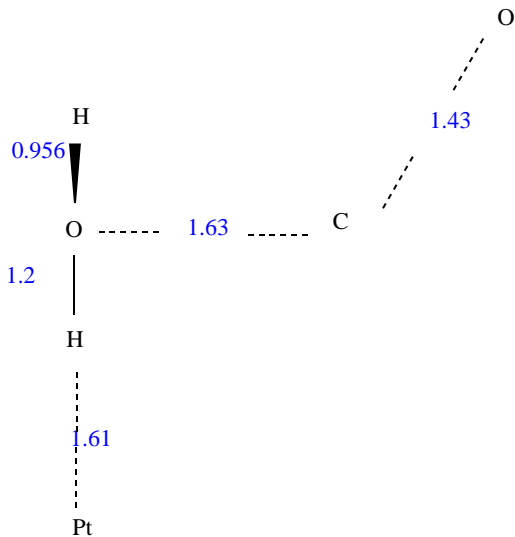}
\caption{Molecular active site in the plane perpendicular to Pt(111) distances in \AA}
\end{figure}

As such, it can be identified by the derivatives with respect to the set of position co-ordinates. There are many of these that are small in magnitude because around the TS there are numerous local minima, where some stabilisation occurs compared to the TS, so it is difficult to obtain a maximum with respect to the reaction path (where the second derivative is negative, since the final slope is ‘downward’ i.e. negative and the initial slope upward and positive). When the values are all small, sign changes are frequent for a small positional change, however these small force constants also imply the influence on total energy is negligible.
The methods of Quantum Monte Carlo calculation are found in detail elsewhere \cite{casino}. The software used is CASINO (well suited to solid state work). We begin by a relatively short Variational Monte Carlo (VMC) step with the input wave-function (plane-waves from embedding the molecular active site wave-function obtained from NECI for TS or asymptotic geometry involving the same atoms for water and CO on Pt(111)) and establish expansion coefficients of a set of polynomials in terms of inter-particle distances (called the Jastrow factor) variationally. The Jastrow factor caters for explicit correlation of electrons, including all three-particle terms, summing over electron pairs and their distance to a given nucleus.
\newpage
Most of the calculation time is consumed by the Diffusion Monte Carlo (DMC) calculation. This takes the correlated input from VMC, in the form of real-space configurations or walkers, distributed to represent electron density and propagates them in an imaginary time-variable obtained by transforming the time-dependent Schr{\"o}dinger equation into a diffusion-drift equation. This stochastic process can be made variational and error in ground-state properties become small in the long-calculation time limit (time is increased in small steps). The accuracy of DMC, which is particularly close to full correlation is second to none.

\section{\label{sec:3} Result.}
For the TS geometry, illustrated in Figure 1, QMC activation energy is 17.0 $\pm$ 0.8 kcal/mol. This low activation energy is readily accessible and lends credibility to a ‘concerted’ reaction mechanism in which O-H elongation in the water molecule occurs simultaneously with a bond-formation between its oxygen atom and the carbon monoxide molecule carbon atom. The leaving hydrogen atom, from water is also forming a Pt-H bond.

{\bf Acknowledgements.}

The QMC calculations were made possible by a PRACE allocation of supercomputer resources: PRACE project 2018184349, allocated 51.6 Million core-hours on the Irene supercomputer (CEA, Bruy\`eres-le-Ch\^atel), near Paris, France.

{\bf Appendix: definition of the periodic cell structure for CASINO input.}

The molecular active site comprises CO adsorbed by the C-atom on a Pt (111) surface atom, and H$_2$O above Pt at the origin, to define the triangle of Pt-atoms in the 111 face.
A Pt slab with 4 atoms in each of 5 layers is defined as face-centred cubic, exposing a compact 111 face. see Figure 2.

\begin{figure}
\includegraphics[width=\textwidth]{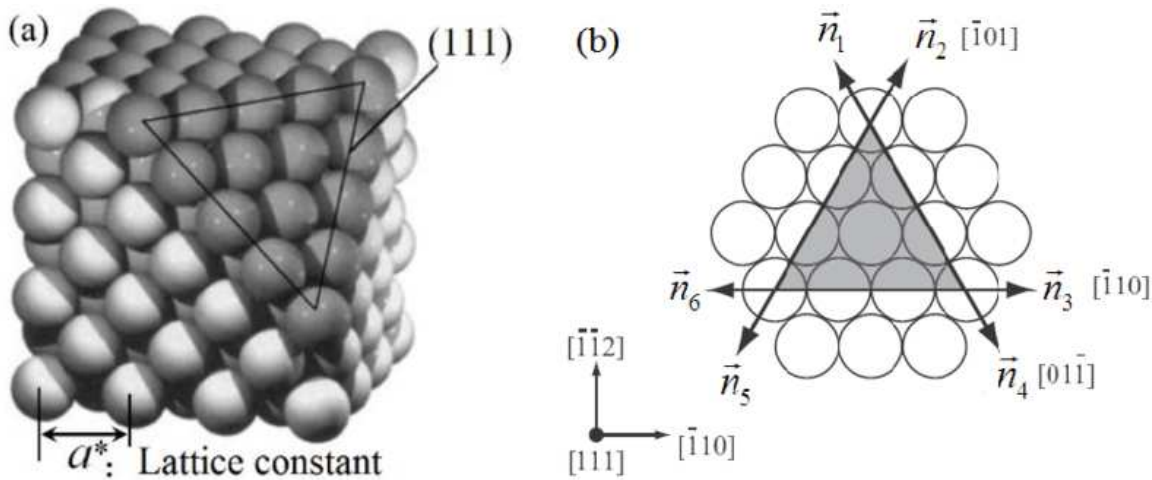}
\caption{Pt slab: 2x2 atoms in a (111) surface, 4 ABAB layers cut from the FCC lattice.\cite{kong}}
\end{figure}

The top 3 layers are spaced according to experimental measurement. The last two derive from the exact bulk a = 3.912 \AA. In plane Pt(111) atom-spacing is always $\sqrt{2} a$ and the last layers are $\sqrt{3} a$ apart to ensure continuity with the bulk. The molecular part is defined with a MRCI wave-function using the Z=60 effective-core potential for Pt-atoms \cite{fpseu} leaving 18 valence electrons per atom. This is embedded in the Pt slab using the method of \cite{bens} for continuity of the electron potential and the whole system wave-function is expanded in plane-waves expressed as B-splines \cite{splin}. Periodicity for ‘in plane, x,y’ directions is physical and on the z-axis the slabs are repeated with a vacuum spacing of 20.8 \AA. The supercell has 1156 electrons. On Irene, 25 DMC steps take 45 mins on 4032 cores.

\end{document}